# Time-dependent droplet detachment behaviour from wettability-engineered fibers during fog harvesting

Arijit Saha, Arkadeep Datta, Arani Mukhopadhyay, Amitava Datta and Ranjan Ganguly[1]

Advanced Materials Research and Applications (AMRA) Laboratory,
Department of Power Engineering, Jadavpur University, Kolkata – 700106, India

**ABSTRACT**

Water collection from natural and industrial fogs has recently been viewed as a viable freshwater source. An interesting outgrowth of the relevant research as focused on arresting of the drift losses (un-evaporated and re-condensed water droplets present in the exhaust plume from industrial cooling towers. Such exploits in fog collection have implemented metal and polyester meshes as fog water collectors (FWC). Fog droplets impinge and deposit on mesh fibers. They coalesce with previously deposited liquid to evolve as larger drops before detaching from the fibers under their own weight, an event largely dependent on the mesh fiber wettability, diameter and its arrangement relative to the fog flow. To better estimate drainage and hence collection from these fibers, the study, focuses on droplet detachment from differently wetted horizontally positioned cylindrical fibers of various diameters, placed orthogonally in the path of an oncoming fog. Droplet detachment volume is found to increase with fiber diameter and fiber surface wettability. Interestingly, in a typical fogging condition, the detachment volume is also found to exhibit a time-dependent behaviour, altering the droplet detachment criteria otherwise predicted from emulation. Our current study sheds light on this unexplored phenomenon.

## 1. INTRODUCTION

Fresh water crisis is one of the most pertinent climate issues of this era, plaguing not only India but the entirety of the globe as seen in recent events. The situation is terrible in countries like India which accounts for only 4% of the world's freshwater resources despite having 16% of the world's population [1]. Nonconventional water sources need to be harnessed to bridge the gap between supply and demand. Fog harvesting is one such water conservation techniques that is looked up with much hope. Fogging represents largely untapped source of water, especially in hilly areas and many industrial zones, the latter primarily referring to harvesting of cooling tower fog. Cooling tower (CT) happens to be one of the prominent sources of industrial fog. For a 500MW power plant, the amount of cooling water required is, 54000 to 60000 $m^3h^{-1}$, from which nearly 3% is lost as vapour and fog from the CT exit.[2] To compensate for this loss, about 900 $m^3h^{-1}$ of make-up water is needed. Tapping even mere 1% of this colossal volume means a saving of 9 tonnes of water in an hour.

In a typical fog harvesting configuration, fog droplets, typically in the size range of 4 – 40 μm, pass through meshes and get captured by the mesh elements. Such meshes or filters are used in large and small-scale atmospheric and industrial water harvesting systems [3,4]. Geometrical structure of the fog collecting mesh and their surface wettability play major roles in the fog collection efficacy. Intuitively, it may appear that a denser net (i.e., having a large fraction of solid fiber per unit projected area of mesh) would imply that more fog particles would collide with the mesh fiber and gather more water. However, this is not typically the case. Smaller void fraction can also have a shielding effect [5], which would lower the mass flow of the fog stream through the mesh itself, and rather divert them around the mesh [6,7,8]. The fraction of the total oncoming fog stream to which the solid fibers pose as obstruction is denoted as aerodynamic efficiency $\eta_{aero}$. The percentage of the fog droplets, whose path ahead of the mesh is geometrically intercepted by the mesh fibers, which actually impinges on the fibers due to inertial impaction, interception and Brownian diffusion is accounted for by the deposition efficiency, or $\eta_{dep}$. It is interesting to note that not all the fog water that deposited on the mesh fiber can be collected. Part of it may be lost due to carryover by the oncoming fog stream, and a part will be lost due to premature dripping [9]. The percentage of water caught by the mesh that drains down the fibers and collects in the water collection manifold at the bottom of the mesh is known as the draining efficiency, or $\eta_{dr}$. The overall collection efficiency, therefore, can be determined by

$$\eta_{collection} = \eta_{capt} \times \eta_{dr} \times \eta_{ac} \quad (1)$$

Commonly deployed FWC meshes consist of fibers woven in orthogonal manner. While the vertical fibers act as a means of sliding of the collected fog water, the horizontal fibers pose challenge to its movement and therefore drainage. An important phenomena in this regard is the dripping or detachment of deposited liquid from the mesh fiber. Our previous studies have focussed on the droplet morphology and detachment criteria from horizontal mesh through numerical simulation [10] and experiments [11], though several important features have further unfolded as our study has progressed. The present work sheds some light on droplet the relatively sparsely characterized phenomenon of droplet detachment from horizontal fiber of varied wettability [12] and diameter, placed in a fog laden flow. Findings of the study are important for designing efficient fog collectors.

## 2. MATERIALS AND METHODS

We developed three types of fibers, superhydrophilic (SHPL), hydrophobic (HPB) and control. For rendering the surface SHPL, a wet chemical etching process was adopted to change the roughness of commercially available aluminum cylindrical fibers (GRADE-6063). The mesh surfaces were first roughened using sandpapers (P220silicon carbide, DV371) to rid it of the oxide protective layer, and then cleansed in acetone and demineralized water respectively using sonication (USB 3.5L H DTC, PCI Analytics) for 15 minutes each. These surfaces were then treated in a solution of 3M HCl aqueous (aq.) solution (MERCK Pvt. Ltd.) for 15 to 20 minutes. This etches the Al surface, creating a micro- and nano-rough texture, rendering the surface SHPL.

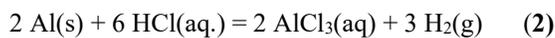

2 Al(s) + 6 HCl(aq.) = 2 AlCl$_3$(aq) + 3 H$_2$(g)    (2)

The SHPL surfaces were next passivated in boiling water for 45 to 60 minutes, which forms nano-böhmite (Al(O)OH) structures on the etched surfaces, improving the surface's resistance to oxidation in the atmosphere, thus making them durable.

To develop hydrophobicity on the mesh surface, the cleansed fibers were dip-coated (NXT dip-KPM, Apex Instruments) (at a coating speed of 3 mm/s) with polydimethylsiloxane (PDMS) (Sigma Aldrich) and cured in an 80°C furnace for 4 hrs.

Stainless steel (SS304 grade Swent feeder needles from Swastik Enterprise$^{TM}$ – India), cleaned in acetone and distilled water, were used as the control fibers.

Sessile-droplet water contact angle on these differently wetted fibers was measured using the methodology defined in [11] using a standard goniometer (Holmarc Opto-Mechatronics Ltd).

A digital screw gauge (Yuzuki$^{TM}$ digital micrometer, least count: 0.001mm) was used to measure each needle's diameter. Before experiment, each fiber was dry cleaned using a blow drier to evaporate any residual fluid and to blow away any dust that might have been deposited on it.

To dispense the fog on the fibers, a medical-grade ultrasonic nebulizer (Yuwell402AI, fog droplet diameter 4-10 μm) was used. The nebulizer's fog-discharge nozzle was precisely positioned and pointed towards the mesh-fiber axis, as shown Fig.1.

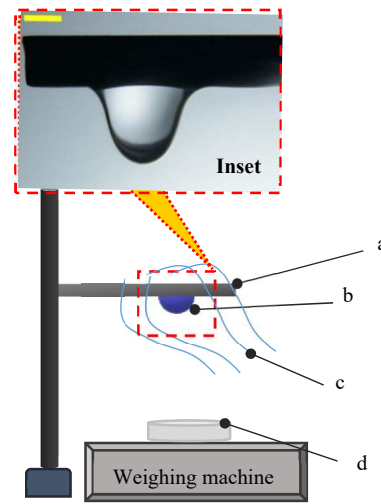

Figure 1: Schematic of the experimental apparatus under the fog condition. Inset shows a typical water droplet pendant upon fog accumulation on the fiber (Scale bar denotes 4mm). Legend: a. fiber, b. water droplet, c. fog-stream, d. collection pot

The distance between the nozzle tip and the needle was fixed and maintained at 2.5 cm. A single-lens digital camera (Nikon D7200 VR with a reversed Nikkon 18-140 mm lens for macro-imaging), was used to record the droplet detachment event. A back-light LED (Osaka Bi-color dimmable led light OS-LED-308) was used at the rear of the droplet to obtain high-quality photos. ImageJ software is used for different measurements pertaining to the images of pendant droplets, e.g., for determine the contact width for the hanging droplet both just before and after detachment events.

## 3. RESULTS AND DISCUSSION

The event of droplet detachment from various wettability-engineered fibers is investigated under fogging conditions.

The fog stream is allowed to strike the horizontally placed fibers when the deposited liquid forms a liquid film, which finally grows into a pendant droplet at the underbelly of the fiber. Eventually the droplet grows to a size that its weight exceeds the adhesion force and it falls down. The process of droplet growth and detachment at the location of fog impingement is noticed. During experiment, the weight of each detached droplet is taken. This revealed an interesting trend. The droplet detachment weight is seen to increase with each detachment event (implying that the size of the detached droplet increases) and then gradually becomes invariant with time (and therefore each detachment event). The variation of detachment weight is shown as a function of droplet number in Fig.2 (A), which clearly points to two distinct regimes; viz the transient and the steady regimes. In the transient regime, the fiber is still 'dry', where the region of fog impaction is still localized. Isolated droplets could be seen on the fiber (except for SHPL fiber) in this regime. This regime gradually shifts to a steady state with gradual coalescence of the micro-droplets into a film, leading to wetting of the fiber; the region of wetting expands and eventually the droplets coalesce into a bigger volume of pendant droplet. These two regimes are depicted in Fig 2(A) on the control fibers of different diameters (regimes are marked as DRY and WET).

This regime shift might be explained as how the fog gradually wets the fiber. In the dry state, the contact line of the droplet is defined as the footprint with which it hangs from the fiber; whereas for the wet state, the contact-line has digressed into taking the smaller drops within its vicinity and therefore increasing its footprint and hence adhesion to the fiber.

From Fig 2(B) we see that the superhydrophilic fibers (SHPL, with CA < 5° measured on a flat plate of similar material and wettability treatment), have the highest detachment weight. The HPB fibers exhibited the least detached droplet weight. In comparison to its "dry" states, the detached droplet weight on wet state marks is just slightly higher (2.93% for 4.36 ± 0.2 mm diameter).

SHPL fiber has the highest surface energy compared to the fibers of other two wettability leading to higher droplet retention capacity.

From figure 2B, we see that he dependence of the droplet detachment weight with fiber diameter tends to taper-off beyond fiber diameter greater than the capillary length of the liquid (2.7mm for water, denoted in the picture as a vertical dashed line). This may be attributed to the fact that the maximum capillary force acting on the droplet scales as the capillary length scale of the droplet [13].

Apart from fog impaction on the hanging droplet, droplet growth is also driven by Laplace pressure, since smaller drops having higher Laplace pressure "pushes" the liquid into the larger hanging drop from its vicinity, since the latter has a lower Laplace pressure due to larger radius of curvature).

Figure 3, denotes such events with pink arrows on control fiber.

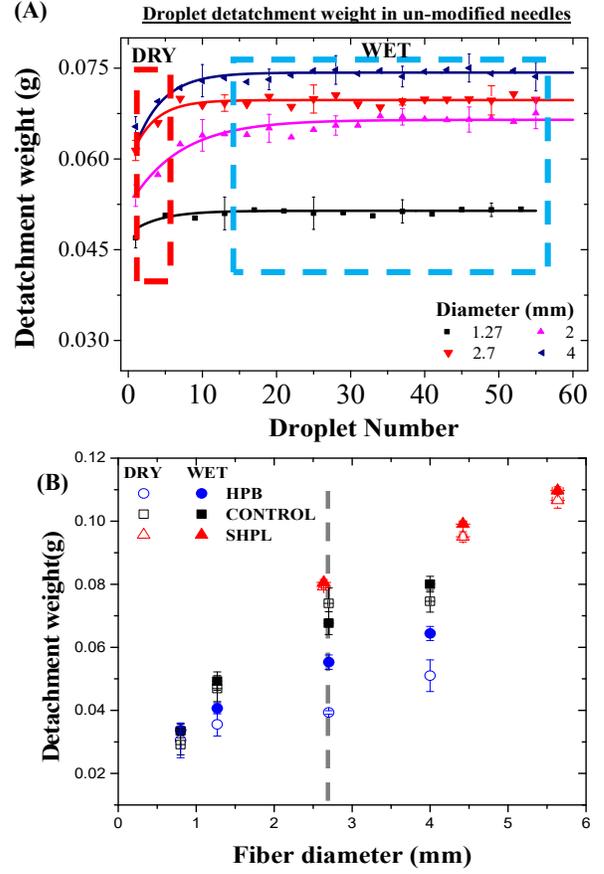

**Figure 2:** (A) Detachment weight mapped with every droplet detachment event (droplet number). The droplet detachment weight initially increases (region marked in red, denoted as DRY) and thereafter becomes invariant (region marked in blue, denoted as WET) with droplet detachment event. Fibers of higher diameter holds more water. Different diameters of control surface (SS304) are used for this study. (B) Comparison of droplet detachment weight from fibers of varied wettability and diameters.

Laplace pressure is the additional pressure a fluid experiences when it is enclosed by a curved surface, and is given by the expression

$$\Delta p = 2\gamma_{LV}\kappa_m + \Delta\rho\vec{g}\cdot\vec{z} \qquad (3)$$

where, $\Delta\rho$ denotes the density difference between the liquid and the gas and $z$ is the local elevation of the surface (from the datum) and the mean curvature $\kappa_m$ can be expressed as [14].

$$\kappa_m = \frac{\nabla F Hess(F)\nabla F^T - |\nabla F|^2 Trace(Hess(F))}{2|\nabla F|^3} \qquad (4)$$

Here, $Hess(F)$ represents the hessian of the surface $F(x,y,z)$ denoting the spatial distribution of the water-air surface [14]. Equation 4 shows that on the same fiber, the droplet with a smaller $\kappa_m$ exerts more excess pressure than that with a larger $\kappa_m$. Therefore, as the smaller droplets were in close vicinity, touching at their bases by a precursor film [15], they would drain into the larger droplet and coalescence. The space freed by these smaller droplets are regenerated with continuous deposition of fog droplets on the fiber. The contact line in the wet state initially increases as a result of this on-going coalescence of the micro-droplets, but thereafter which it attains a steady state, characterized by the invariance towards droplet detachment event (droplet number). This translates to the trend as depicted in Fig. 2(A) for fibers of different diameter.

The variation of the detachment weight with fiber radius for dry and wet conditions on fibers of three different wettability is plotted in Fig 2(B). The surface energy difference is what causes this difference in the detachment volume for different wettability fibers. In comparison to hydrophobic surfaces, hydrophilic surfaces have higher surface energy. A surface with more surface energy can accommodate larger droplets on it. The surface pinning force is another crucial factor that affects the detachment weight. The pinning force effect also rises with fiber diameter. The fiber can hold more water droplets when there is more pinning force acting along its three-phase contact line. As the HPB (PDMS-coated) surface has less surface energy and a weaker pining force (characterized by a lower contact angle hysteresis), it is expected to hold smaller droplets before detachment. Figure 2(B) also shows the difference between dry state and wet state detachment weight. For the SHPL fiber, the detachment weight curves for dry and regimes almost overlap with each other because the droplet detachment from the horizontal fiber takes place from a water film on the surface for both the dry and wet states. However, when compared to the other two wettability fibers, HPB fiber exhibits the greatest variation, because it resists the formation of a water film on it in both dry and wet states. Hoever, but in the wet state, small droplets are found in the vicinity of the large droplet, which collides with the large droplet at the time of detachment due to Laplace pressure. This phenomenon creates the difference between dry and wet state detachment weight.

Another important parameter for determining the variation of detached droplet weight is droplet-fiber adhesive force, which is a function of the liquid surface tension and the contact perimeter that the droplet makes with the fiber. The adhesive force balances the hanging droplet's weight. In contrast to the other two wettability fibers, the SHPL fiber-droplet contact line is maximum since its pinning force effect is maximum as seen Fig. 3. This is observed in both the dry and the wet states. The droplet-fiber contact length, $L_c$ for HPB fiber is particularly low because of its low surface energy, which prevents the droplet from expanding its perimeters on the fiber. The contact footprint (or the contact length, $L_c$) of the hanging droplet seems to be a major contributing factor for the trend we see in Fig. 2(B).

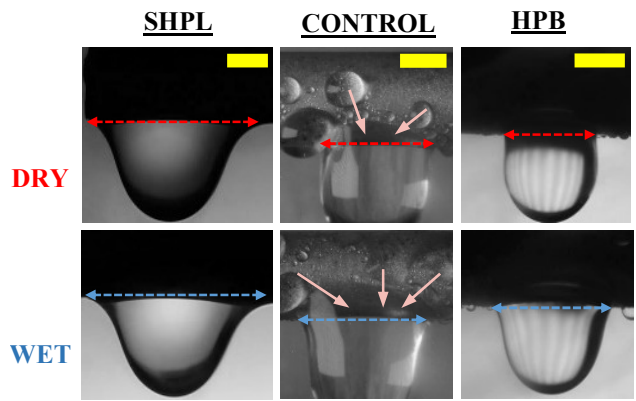

**Figure 3**: Droplet detachment from three wettability fibers in both the dry and the wet states, captured just before detachment. Contact length ($L_c$) is demarcated as [← – →] and [← – →] in DRY and WET states respectively. The arrows in pink denotes Laplace pressure driven flow from smaller drops from in and around the much larger hanging drop.
**Scale bar denotes 2mm.**

Experimentally, it is quite difficult to explicitly gauge the contact line of the hanging water droplet on a horizontally placed fiber. The contact width, marked by horizontal arrows in Fig.3 (Red and blue for DRY and WET state respectively) could be taken as an approximation of the contact line, considering the length scale. The maximal width of the growing droplet is measured in both the events. This provides a rough estimate of the detachment weights, given the practical constraints on the pendant droplet imposed by the solid surface geometry and contact line conditions (i.e., a moving or a pinned contact line) [16,17].

With successive deposition of fog, there is a gradual increase in contact width with fiber diameter for both the regimes, taken just prior to detachment phenomena. This is attributed to the larger footprints of the pendant drop that the fiber can accommodate. The contact width lies in the two extremes for the SHPL and HPB fibers as shown in Fig. 4, with the general trend remaining the same as discussed for control fibers. The contact width differs between the wet and dry states in HPB (18 % -13.46%) and the SHPL (3 % - 0.5%) cases and is quite smaller than that for the control fibers. For the SHPL fibers, this could be attributed to it having a higher surface energy. It being super hydrophilic, droplet spreads on its surface forming a film of water on the fiber surface, raising the upper limit of contact width. This also attributes to it holding a much higher volume of water during detachment. The formation of this liquid film also explains the overlapping of the contact width data in Fig. 4 (in both the wet and dry regimes) for SHPL cases.

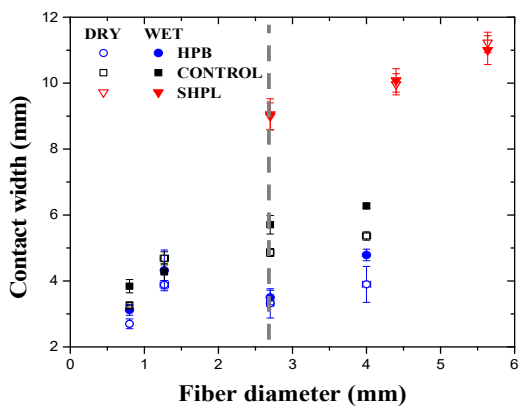

**Figure 4** Variation of contact width of the hanging droplet from differently wettability engineered fibers of varied diameters just before detachment. Variation of droplet detachment weight with fiber diameter tends to taper-off beyond fiber diameter greater than the capillary length of the liquid (2.7mm for water, shown by the vertical dotted line).

There is also no distinct 'dry' and 'wet' state for the SHPL cases, due to this liquid film formation. This also explains the overlapping of detachment weight data in 'dry' and 'wet' regimes (Fig 2B) and high detachment weight for SHPL cases. The observations the HPB fibers were similar to those made for SHPL fibers, with only one major difference; the contact width for HPB fibers were at the lowest end for the spectrum. This is attributed to its low surface energy, which restricts water to adhere to the surface and inhibits any film formation. This is also supported by the detachment weight data for both dry and wet regimes (Fig. 2B).

## 4. CONCLUSIONS

Liquid droplet detachment from horizontal fibers of varied diameter and varied wettability is studied experimentally in the context of fog harvesting. Liquid deposition is carried out through fog impaction. The role of fiber wettability – superhydrophilic (SHPL), control and hydrophobic (HPB) – on droplet detachment dynamics is explored. It is observed that the SHPL fibers can hold the largest droplets before detachment, followed by the control and HPB fibers, along with the general trend suggesting a higher droplet detachment weight with increase in fiber radius. The dependence of the droplet detachment weight with fiber diameter tends to taper-off beyond fiber diameter greater than the capillary length of the liquid (2.7mm for water). It is observed that the droplet contact width on the SHPL fibers is more in comparison with the control and HPB fibers, which also supports droplet detached weight trends and is attributed to the balancing act of the adhesive surface tension force to the weight of the pendant droplet.

**ACKNOWLEDGEMENTS**

The authors gratefully acknowledge the funding from DST SERB (Grant No: CRG/2019/005887).

**REFERENCES**

[1] http://mospi.nic.in/sites/default/files/Statistical_year_book_india_chapters/ch2.pdf [Accessed on 05/07/2022]
[2] Ghosh, Ritwick, Tapan K. Ray, and Ranjan Ganguly. "Cooling tower fog harvesting in power plants–A pilot study." Energy 89 (2015): 1018-1028.
[3] Chayan Das, Ritwick Ghosh, Amitava Datta, Ranjan Ganguly, "Wettability engineering: A skin deep approach of solving the energy-water nexus" Begell House Inc. Book title: Advances in Multiphase Flows. ISBN: 978-1-56700-504-2.
[4] Damak, M., & Varanasi, K. K. (2018). Electrostatically driven fog collection using space charge injection. Science advances, 4(6), eaao5323.
[5] de Dios Rivera, J. (2011). Aerodynamic collection efficiency of fog water collectors. Atmospheric Research, 102(3), 335-342.
[6] Park, Kyoo-Chul, et al. "Optimal design of permeable fiber network structures for fog harvesting." Langmuir 29.43 (2013): 13269-13277.
[7] Shi, Weiwei, et al. "Fog harvesting with harps." ACS applied materials & interfaces 10.14 (2018): 11979-11986.
[8] Regalado, Carlos M., and Axel Ritter. "The design of an optimal fog water collector: A theoretical analysis." Atmospheric Research 178 (2016): 45-54.
[9] Ghosh R, Ganguly R. Fog harvesting from cooling towers using metal mesh: Effects of aerodynamic, deposition, and drainage efficiencies. Proceedings of the Institution of Mechanical Engineers, Part A: Journal of Power and Energy. 2020;234(7):994-1014.
[10] Arani Mukhopadhyay, Partha Sarathi Dutta, Amitava Datta, and Ranjan Ganguly, Liquid droplet morphology on the fiber of a fog harvester mesh and the droplet detachment conditions under gravity, Proceedings of the 8th International and 47th National Conference on Fluid Mechanics and Fluid Power (FMFP), December 09-11, 2020, IIT Guwahati, Guwahati-781039, Assam, India
[11] Arkadeep Datta, Arani Mukhopadhyay, Partha Sarathi Dutta, Arijit Saha, Amitava Datta and Ranjan Ganguly, Droplet detachment from a horizontal fiber of a fog harvesting mesh, FMFP2021–190, Proceedings of the 48thNational Conference on Fluid Mechanics and Fluid Power (FMFP), December 27-29, 2021, BITS Pilani, Pilani Campus, RJ, India.
[12] De Gennes, Pierre-Gilles, Françoise Brochard-Wyart, and David Quéré. Capillarity and wetting phenomena: drops,

bubbles, pearls, waves. Vol. 315. New York: Springer, 2004.

[13] Lorenceau, É., Clanet, C., & Quéré, D. (2004). Capturing drops with a thin fiber. Journal of Colloid and Interface Science, 279(1), 192–197.

[14] Goldman, R., 2005. Curvature formulas for implicit curves and surfaces. Computer Aided Geometric Design, 22(7), pp.632-658.

15 Bonn, D.; Eggers, J.; Indekeu, J.; Meunier, J.; Rolley, E. Wetting and Spreading. Rev. Mod. Phys. 2009, 81, 739−805.

[16] Kumar, A., Gunjan, M. R., & Raj, R. (2020). On the validity of force balance models for predicting gravity-induced detachment of pendant drops and bubbles. Physics of Fluids, 32(10), 101703.

[17] Kumar, A., Gunjan, M. R., Jakhar, K., Thakur, A., & Raj, R. (2020). Unified Framework for Mapping Shape and Stability of Pendant Drops Including the Effect of Contact Angle Hysteresis. Colloids and Surfaces A: Physicochemical and Engineering Aspects, 124619.